\documentclass[nofootinbib,preprintnumbers,prd,superscriptaddress,showpacs,twocolumn]{revtex4}

\usepackage[titletoc]{appendix}
\usepackage{color}
\usepackage{hyperref}

\usepackage[rightcaption]{sidecap}
\usepackage{subfigure}
\usepackage{comment}

\usepackage{graphicx}
\usepackage{amsmath}
\usepackage{amsfonts}
\usepackage{amsthm}
\usepackage{mathrsfs}
\usepackage{amssymb}
\usepackage{epsfig}
\usepackage{amscd}
\usepackage{dcolumn}
\usepackage{bm}
\usepackage{natbib}
\usepackage{url}
\usepackage{xspace}

\usepackage{dcolumn}

\usepackage{array}
\usepackage{ctable}
\usepackage{multirow}
\usepackage{siunitx}
\usepackage{longtable}
\usepackage{tabularx}
\usepackage{booktabs}

\usepackage{amsthm}
\usepackage{mathrsfs}

\usepackage{epsfig}
\usepackage{amscd}

\usepackage{bm}
\usepackage{natbib}
\usepackage{url}
\usepackage{xspace}
\usepackage{kantlipsum}

\graphicspath{{Graphics/}}

\def\be{\begin{equation}}
\def\ee{\end{equation}}
\def\bea{\begin{eqnarray}}
\def\eea{\end{eqnarray}}

\def\prd{Phys. Rev. D}

\def\mnras{MNRAS}

\def\apj{ApJ}
\def\apjl{ApJ Lett.}

\def\aap{A\&A}

\definecolor{vividviolet}{rgb}{0.62, 0.0, 1.0}
\definecolor{amaranth}{rgb}{0.9, 0.17, 0.31}
\definecolor{palatinateblue}{rgb}{0.15, 0.23, 0.89}
\definecolor{brightpink}{rgb}{1.0, 0.0, 0.5}
\definecolor{cornflowerblue}{rgb}{0.39, 0.58, 0.93}
\definecolor{deepcarminepink}{rgb}{0.94, 0.19, 0.22}
\definecolor{radicalred}{rgb}{1.0, 0.21, 0.37}

\hypersetup{ linktoc=all,
    colorlinks, linkcolor={palatinateblue},
    citecolor={brightpink}, urlcolor={amaranth}
}

\begin{document}

\title{Addressing the $H_0$ tension through matter with pressure and no early dark energy}

\author{Youri Carloni}
\email{youri.carloni@unicam.it}
\affiliation{Universit\`a di Camerino, Via Madonna delle Carceri, Camerino, 62032, Italy.}
\affiliation{INAF - Osservatorio Astronomico di Brera, Milano, Italy.}
\affiliation{Istituto Nazionale di Fisica Nucleare, Sezione di Perugia, Perugia, 06123, Italy.}

\author{Orlando Luongo}
\email{orlando.luongo@unicam.it}
\affiliation{Universit\`a di Camerino, Via Madonna delle Carceri, Camerino, 62032, Italy.}
\affiliation{INAF - Osservatorio Astronomico di Brera, Milano, Italy.}
\affiliation{Istituto Nazionale di Fisica Nucleare, Sezione di Perugia, Perugia, 06123, Italy.}
\affiliation{SUNY Polytechnic Institute, 13502 Utica, New York, USA.}
\affiliation{Al-Farabi Kazakh National University, Al-Farabi av. 71, 050040 Almaty, Kazakhstan.}

\author{Marco Muccino}
\email{marco.muccino@unicam.it}
\affiliation{Universit\`a di Camerino, Via Madonna delle Carceri, Camerino, 62032, Italy.}
\affiliation{Al-Farabi Kazakh National University, Al-Farabi av. 71, 050040 Almaty, Kazakhstan.}
\affiliation{ICRANet, Piazza della Repubblica 10, 65122 Pescara, Italy.}

\begin{abstract}
We propose that the Hubble tension arises due to an unaccounted additional component, that behaves as \emph{matter with pressure}. We demonstrate that this fluid remains subdominant compared to both dust and radiation throughout nearly the entire universe expansion history. Specifically, the additional fluid satisfies the Zel'dovic limit with a constant equation of state, $\omega_s > 0$, and a quite small normalized energy density, $\Omega_s$. Accordingly, this component modifies both the sound horizon and the background expansion rate, \emph{acting quite differently from early dark energy models}, without significantly affecting the other cosmological parameters. To show this, we perform a Monte Carlo Markov chain analysis of our model, hereafter dubbed $\Lambda_{\omega_s}$CDM paradigm, using the publicly available \texttt{CLASS} Boltzmann code. Our results confirm the presence of this fluid, with properties that closely resemble those of radiation. We find best-fit values that satisfy $\omega_s \lesssim \omega_\gamma$ and a relative energy density $\Omega_s / \Omega_\gamma = 0.45$, with $\omega_r$ and $\Omega_r$ the equation of state and density of photons, respectively. The effective fluid may be associated with generalized K-essence models or, alternatively, with Proca-type vector fields, albeit we do not exclude \emph{a priori} more exotic possibilities, i.e., dark radiation, axions, and so on. Physical implications of our results are analyzed in detail, indicating a statistical preference for the $\Lambda_{\omega_s}$CDM scenario over the conventional $\Lambda$CDM background.
\end{abstract}


\maketitle
\tableofcontents

\section{Introduction}
\label{intro}

Cosmology is experiencing a phase of heat debate, as due to the impressive new amount of data points, certifying unexpected dark energy evolution within the cosmic background \cite{DESI:2025zgx}\footnote{Strong criticisms to these outcomes can be found in Refs.~\cite{Carloni:2024zpl,Luongo:2024fww,Alfano:2024fzv}.} and seeking alternative dark matter particle candidates following the failure of weakly interactive massive particle (WIMP) hypothesis\footnote{Current bounds show that ultralight or supermassive fields could be suitable candidates, excluding \emph{de facto} WIMPs \cite{Bertone:2018krk}. Alternatives may be extensions Einstein's gravity \cite{Capozziello:2019cav} or spacetime solutions, within general relativity, behaving as particles \cite{Luongo:2025iqq}.}.

In this \emph{mare magnum} of challenges for modern cosmology, the perplexing \emph{Hubble tension} refers to the discrepancy between local measurements and values inferred from early-universe observations of the present-day expansion rate of the universe, i.e., the Hubble constant $H_0$.
More precisely, this tension arises when comparing local measurements by the SH0ES collaboration that, utilizing Cepheid-calibrated cosmic distance ladder, reports $H_0^{\rm R}=(73.04\pm1.04)$~km/s/Mpc \cite{2022ApJ...934L...7R} or cosmographic methods \cite{Aviles:2012ay,Dunsby:2015ers,Luongo:2020aqw,Luongo:2012dv,Bamba:2012cp}, and values obtained from the cosmic microwave background (CMB), within the  $\Lambda$CDM background, leading to $H_0^{\rm P}=(67.36\pm0.54)$~km/s/Mpc \cite{Planck2018}.
This tension at $4.1\sigma$ confidence level\footnote{Given that this discrepancy is corroborated by multiple independent probes, it is unlikely that a single systematic error accounts for the full disagreement, as above stated.} appears quite unexpected and may hint at new physics \cite{Cortes:2023dij,Hu:2024big,Hu:2023jqc,Gomez-Valent:2024tdb,Yang:2019jwn,Pan:2020bur}, i.e., new fields \cite{Niedermann:2020dwg,Chatrchyan:2024xjj}, new dark interactions \cite{Blinov:2020uvz,Nygaard:2023gel,Mirpoorian:2024fka,Odintsov:2020qzd}, or just issues related to the experimental procedures \cite{Pedrotti:2024kpn,Lee:2025yah,Banik:2025dlo}. For a review see e.g. Ref.~\cite{Vagnozzi:2023nrq}.

To address the persistent $H_0$ tension, we may focus on two main avenues. The first is relaxing the assumption that Milky Way dust properties are well-defined at low redshifts, thereby allowing for spatial or temporal variations in dust composition and extinction laws, at $ z \sim 0 $. Alternatively, assuming that type Ia supernovae (SNe~Ia) remain reliable and high-precision standardizable candles, a second possibility to resolve the Hubble tension may lie in early-universe physics, potentially requiring modifications to the pre-recombination dynamics or extensions to the standard $\Lambda$CDM framework, at redshifts before the CMB.


Hence, approaches to alleviate the $H_0$ tension are broadly classified into \emph{late- and early-time} scenarios. Practically speaking, provided that the angular size of the sound horizon $\theta_\star = r_\star/[(1+z_\star)D_A(z_\star)]$ at the recombination\footnote{In this work, we treat the redshifts of recombination, photon decoupling, and last scattering as interchangeable, following the convention adopted in several studies \cite{Hu:2001bc,Desjacques:2015yfa,Planck:2018vyg,Kamionkowski:2022pkx,Lynch:2024gmp,Mirpoorian:2024fka,Kou:2024rvn}.} redshift $z_\star$ is measured with a precision of $0.03\%$ \cite{Planck2018}, late-time solutions alter the angular diameter distance $D_A(z_\star)$ of the last scattering surface, whereas early-time solutions typically modify the comoving sound horizon $r_\star = r_s(z_\star)$, furnishing the options below.
\begin{itemize}
    \item[-] \textbf{Late-times}. Modifying $H(z)$ between $z=0$ and $z_\star$ implies a different value of $H_0$, keeping the comoving sound horizon $r_\star$ and the angular diameter distance $D_A(z_\star)$ fixed. This ensures that the angular scale of the sound horizon, $\theta_\star$, remains unchanged.
    \item[-] \textbf{Early-times}. Modifying $r_\star$, by changing either $z_{\star}$, the sound of speed, $c_s(z)$, or the Hubble parameter, $H(z)$, \emph{always before recombination}. Accordingly, $H_0$, as embedded in the expression of $D_A$, varies to preserve the angular scale of the sound horizon $\theta_\star$.
\end{itemize}

The late-time approach appears quite disfavored, since $H(z)$ in that regime is well-constrained and it turns out to be difficult to expect physical modifications. Conversely, typical early-time modifications tend to reduce $r_\star$, reconciling \emph{de facto} the locally measured value of $H_0$ with CMB observations, by keeping $\theta_\star$ fixed \cite{Schoneberg:2021qvd,Kamionkowski:2022pkx, Poulin:2023lkg}.

Motivated by the above considerations, we focus here on early-times and investigate an alternative approach to early dark energy (EDE) models, without involving scalar fields. To do so, we address the $H_0$ tension \emph{predicting the existence of an additional barotropic fluid} that satisfies the Zel'dovic limit, namely \emph{matter with pressure}, that exhibits a positive equation of state (EoS), different from radiation and EDE models. In so doing, we directly modify the comoving sound horizon, $r_s$, and $H(z)$, as due to the presence of this additional fluid \emph{before recombination}. This procedure leads to an effective increase in the Hubble constant, slightly modifying the $\Lambda$CDM model into a fast and direct extension. In particular, the additional barotropic fluid turns out to be subdominant with respect to dust, neutrinos, radiation and so on. To show this, we compute exclusion plots, where the range of values for the fluid magnitude and EoS is carefully limited, showing to be substantially smaller than radiation, i.e., \emph{de facto} not substantially modifying the CMB structure as well. Afterwards, once fixed suitable priors, we perform a cosmological fit, including neutrinos and radiation, in fulfillment with current Planck's results. The fit is worked out by analyzing our so-constructed paradigm, hereafter dubbed $\Lambda_{\omega_s}$CDM model, by means of a Monte Carlo Markov chain (MCMC), modifying the free-available \textit{CLASS} code\footnote{\url{http://class-code.net/}} \cite{2011arXiv1104.2932L}, employing the ranges of values that cannot exclude the presence of a further fluid. Suitable findings, in a spatially flat universe, are thus preliminary obtained, indicating that our slight extension of the $\Lambda$CDM model is not \emph{a priori} disfavored with respect to the usual case. We find inconsistency with a purely-behaved stiff matter fluid, but rather with an EoS smaller than that of radiation, i.e., $\omega_s\lesssim 1/3$, with a normalized density, quite smaller than that of radiation. Physical consequences and corresponding explanations toward this matter-like fluid, inspired by pure stiff matter are thus explored in detail. Further, its fundamental origins,  in terms of extra fields, are summarized.

The paper is structured as follows. In Sect.~\ref{sezione2}, we discuss the Hubble tension, summarizing the physical approaches and explaining our new treatment. In Sect.~\ref{sezione3}, we work out our analysis, based on the use of a MCMC simulation and discuss the corresponding outcomes. Accordingly, we physically justify our fluid and the physics behind it. In Sect.~\ref{sezione4}, we conclude our work, emphasizing consequences and perspectives of our treatment.

\section{Tackling the Hubble tension}\label{sezione2}

Alleviating the $H_0$ tension involves keeping the angular scale $\theta_\star$ fixed while modifying the comoving sound horizon $r_s$ or the angular diameter distance $D_A$, as previously discussed. In particular, by considering the standard approach, the value of $H_0$ is effectively embedded within the calculation of $r_\star=r_s(z_\star)$, given by
\begin{equation}
\label{eq:0}
r_\star = \int_{z_{\star}}^{\infty} \frac{c_s(z)}{H(z)}dz = \frac{c}{\sqrt{3}H_{\star}} \int_{z_{\star}}^{\infty} \sqrt{\frac{\rho_\star}{\rho(z)}}\frac{dz}{\sqrt{1 + R}},
\end{equation}
where $H_{\star}=H(z_\star)$ and $\rho_\star=\rho(z_\star)$ are the Hubble parameter and the total density $\rho(z)$ at the recombination, respectively, and $R=(3/4)(\Omega_b/\Omega_\gamma)(1 + z)^{-1}$. The quantities $\Omega_{b}$ and $\Omega_{\gamma}$ are the present-day (normalized to the critical density) baryon and photon densities.

Hence, to maintain $\theta_\star$ fixed, if $r_\star$ decreases then $D_{\rm A}(z_\star)$ might decrease. The latter is given by
\begin{equation}
D_{\rm A}(z_\star) = \frac{c}{1+z}\int_0^{z_\star}\frac{dz}{H(z)},
\end{equation}
leading to a ``decrease'' of $H_0$,  inferred from local measurements. As late-time constraints disfavor changes in $H(z)$ after recombination, we focus on early-time solutions, outlined in the following three classes of approaches, plus a fourth scenario that consists of our new fluid, suggesting a minimal corrected $\Lambda_{\omega_s}$CDM paradigm, inspired by matter with pressure.

\subsection{Modifying the sound speed}

The first solution faces how to change the sound speed. Thus, before recombination, photons and baryons are tightly coupled through Thomson scattering and behave effectively as a \emph{single fluid}. The corresponding adiabatic sound speed (i.e., at constant entropy $S$) definition is
\begin{equation}
    c^{2}_{s} = c^{2}\left( \frac{\partial P}{\partial \rho} \right)_{S}.
\end{equation}
As the \emph{baryons are assumed to be non-relativistic} this photon-baryon fluid exhibits a total pressure due to the photon component only, $P=\rho_\gamma/3$, and a total energy density given by the sum of the two components, yielding $\rho = \rho_\gamma + \rho_b$.
Thus, the sound speed is given by
\begin{equation}
    c_s^2 = \frac{c^2}{3(1 + R)}.
\end{equation}
It is clear that altering this value leads to a modification of the comoving sound horizon. Specifically, if we decrease it by introducing new additional species into the plasma, the comoving sound horizon becomes smaller, increasing the inferred value of $H_0$ \cite{Evslin:2017qdn}.

\subsection{Varying the recombination redshift}

The second possibility to heal the $H_0$ tension at early-times is to vary the recombination redshift. In particular, changes in the recombination history directly impact the size of the sound horizon in the photon–baryon plasma. However, since the corresponding angular scale at decoupling, $\theta_\star$, is precisely measured, any variation in its physical scale must be balanced by a corresponding change in the angular diameter distance to the surface of last scattering. The CMB’s sensitivity to the ionization history depends upon the photon scattering rate, which is governed by the ionization fraction,
\begin{equation}
    X_e(z) = \frac{n_e(z)}{n_H(z)},
\end{equation}
where $n_e$ and $n_H$ denote the number densities of free electrons and hydrogen nuclei, respectively. The recombination redshift is typically defined by the condition $X_e(z_\star) = 0.5$. A modified recombination history leads to a different ionization fraction, thereby shifting $z_\star$ and altering $r_\star$ \cite{Lynch:2024hzh,Lynch:2024gmp, Mirpoorian:2024fka}. In particular, a lower $z_\star$ reduces $r_s$, which increases the value of $H_0$.

\subsection{Resorting early dark energy}

A third and very promis approach to alleviating the $H_0$ tension involves the presence of EDE. This term refers to any form of dark energy that becomes dynamically relevant before recombination, thus modifying the Hubble expansion rate and reducing the comoving sound horizon. The presence of a dark energy component with a time-dependent EoS, non-negligible at early times, has been extensively studied \cite{Doran:2006kp, Linder:2008nq, Grossi:2008xh, Maggiore:2011hw, Pettorino:2013ia, Poulin:2018cxd,Niedermann:2019olb, Braglia:2020bym,Smith:2020rxx,Gomez-Valent:2021cbe,Sabla:2022xzj}, originally to investigate its impact on structure formation \cite{Doran:2001rw, Doran:2002ec, Wetterich:2004pv}, and its effect on increasing the value of $H_0$ depends on the specific model considered. The impact of EDE is determined by the fractional contribution of the field to the total energy density, expressed as $f_{\rm EDE}(z)=\rho_{\rm EDE}(z)/\rho(z)$, which increases over time. In particular, the scalar field initially remains frozen in the potential, resulting in a nearly constant energy density. Later, it is released from this configuration by a physical mechanism, such as a drop in Hubble friction below a critical threshold or a phase transition that alters the shape of the potential. Once the field becomes dynamical, its energy density dilutes faster than that of matter, leading to a rapid decay of its contribution to the total energy budget.

Therefore, the impact of EDE on the expansion rate is localized in redshift, occurring prior to recombination, and becomes negligible at late times. This allows it to reduce the comoving sound horizon, leaving the angular diameter distance unaltered.

A successful alleviation of $H_0$ tension generally requires a fractional contribution of $f_{\mathrm{EDE}} \sim 10\%$, peaking at redshifts $z \sim 10^3$–$10^4$, and subsequently diluting at a rate at least as fast as that of radiation \cite{Kamionkowski:2022pkx, Poulin:2023lkg}.

\subsection{Introducing matter with pressure}

As fourth attempt, we here propose a novel early-time resolution to the Hubble tension by introducing a new barotropic fluid, i.e., a pseudo-relativistic matter characterized by an arbitrary EoS, which reduces the comoving sound horizon, changing at the same time $H(z)$, and that increases the inferred value of the Hubble constant.

In this way, our new setup is determined by a tightly coupled fluid composed of three main components:
\begin{itemize}
    \item[-] Photons, with $ P_\gamma = \rho_\gamma/3$.
    \item[-] Baryons, with negligible pressure, $P_b \approx 0 $.
    \item[-] Relativistic matter, with $ P_s = \omega_s \rho_s $, where $\omega_s>0$ and not necessarily $\omega_s=1$ (stiff matter).
\end{itemize}

\begin{table*}[ht]
\centering
\renewcommand{\arraystretch}{1.}
\begin{tabular}{p{8.cm} p{8.cm}@{}}
\hline\hline
\multicolumn{1}{c}{\textbf{Matter with pressure}} & \multicolumn{1}{c}{\textbf{Early dark energy}} \\
\hline
\multicolumn{2}{c}{\emph{{\bf Thermodynamics}}}\\
\begin{enumerate}
    \item Constant and positive $\omega_s$, satisfying the Zel'dovic limit. Direct thermodynamic interpretation.
    \item $\rho_s(z) \propto (1+z)^{3(1 + \omega_s)}$, not designed to decay.
\end{enumerate} &
\begin{enumerate}
    \item Time-varying EoS. Nearly constant density.
    \item Both $\omega_{\rm EDE}$ and $\rho_{\rm EDE}$ are designed to decay after recombination.
\end{enumerate} \\
\multicolumn{2}{c}{\emph{{\bf Physical origin and properties}}} \\
\begin{enumerate}
    \item Hypothetical matter or radiation component with pressure. Elastic properties.
    \item Always present but generally subdominant, affecting early universe expansion.
\end{enumerate} &
\begin{enumerate}
    \item Typically scalar fields, e.g., axion-like. It requires an additional non-barotropic constituent.
    \item Fine-tuned: designed to dominate only during a brief early epoch, and to vanish later.
\end{enumerate} \\
\multicolumn{2}{c}{\emph{{\bf Observable effects}}}\\
\begin{enumerate}
    \item It mainly modifies early background expansion. \emph{Directly detectable}.
    \item It affects primordial evolution. Diluted at late times.
\end{enumerate} &
\begin{enumerate}
    \item It might alter recombination history only. \emph{Directly undetectable}.
    \item Designed to alleviate specific cosmological tensions by turning off dynamically.
\end{enumerate}\\
\hline\hline
\end{tabular}
\caption{Comparison between a positive-$\omega_s$ barotropic fluid (interpreted as matter with pressure) and EDE paradigms.
}
\label{tab:comparison}
\end{table*}

Here, the total energy density and pressure are
\begin{subequations}
\begin{align}
    &\rho = \rho_\gamma + \rho_b + \rho_s,\\
    &P = P_\gamma + P_s = \frac{1}{3} \rho_\gamma + \omega_s \rho_s,
\end{align}
\end{subequations}
and the adiabatic sound speed becomes
\begin{equation}
    c^{2}_{s} = c^{2}\left( \frac{\partial P}{\partial \rho} \right)_S = \frac{c^{2}\left(\partial \rho_\gamma + 3 \omega_s \partial \rho_s\right)}{3\left(\partial \rho_\gamma + \partial \rho_b + \partial \rho_s\right)}.
\end{equation}
Introducing a new dimensionless ratio involving the new fluid and the photon densities,
$W \equiv 3(1+\omega_s)\rho_s/(4\rho_\gamma)$, the modified sound speed before the recombination is
\begin{equation}
\label{eq:cszel}
    c_s^2 = \frac{c^2(1+3\omega_s W)}{3(1+R+W)},
\end{equation}
\emph{indicating that this new species alters not only the Hubble parameter, but also the primordial plasma}.

Barotropic fluids differ from a scalar field as the kinetic energy of the latter may be either positive or negative, see e.g. \cite{Linder:2008ya,Steinhardt:1999nw, Chiba:2005tj, Scherrer:2007pu, Scherrer:2008be, Wolf:2023uno, Carloni:2024ybx}.
Ensuring the Zel'dovic limit, namely $\omega_s > 0$, we end up with the following requirements.
\begin{itemize}
\item[-] The fluid has \emph{positive pressure}, say it may contribute as an attractive component to the stress-energy tensor, contrasting any pure dark energy term \cite{2018PhRvD..98j3520L, 2022CQGra..39s5014D}.
\item[-] The fluid dilutes faster than cold dark matter and baryons as the universe expands\footnote{For the intriguing case of \emph{dark matter}, instead of pure matter, with pressure, refer to as Refs. \cite{Faber:2005xc,Dunsby:2016lkw,Luongo:2022uva,Dunsby:2023qpb,Luongo:2025iqq}.}, providing an energy density $
\rho_s(z) \propto (1+z)^{3(1 + \omega_s)}$.
\item[-] Physically, it would contribute non-negligibly at early times to alter $H_0$, remaining quite subdominant at all stages of universe's evolution.
\end{itemize}

Our proposed class of barotropic fluids represents a naive and classical example in theoretical cosmology, since it simply guarantees the Zel'dovich limit. Besides pure radiation, corresponding to ultrarelativistic particles, such as photons, neutrinos, etc., our $\Lambda_{\omega_S}$CDM model is inspired by the possible existence of genuine stiff matter, characterized by $\omega_{sm} = 1$ \cite{Chavanis:2014lra, Gron:2024bur}.

However, quite more interestingly for our purposes, the case $0 < \omega_s < 1/3$ is effectively explored here, see Fig. \ref{fig:Osws}. Indeed, even though this scenario has not been directly observed yet, in principle it appears quite possible, as it corresponds to a matter that has a non-zero pressure, being pseudo-relativistic, but subdominant to radiation, see Fig. \ref{fig:eq}.

Thus, if this fluid exists, one can wonder which differences may occur between this approach and a pure EDE contribution. To this end, the key differences between a positive-$\omega$ barotropic fluid and EDE are summarized in Table~\ref{tab:comparison}. In summary, even though both a barotropic fluid, with $\omega_s > 0$, and EDE can impact the early universe, \emph{they are fundamentally different for their origin and behavior}. Mainly, a barotropic fluid follows a constant EoS and scales predictably with the expansion, being subdominant, albeit always present along the universe evolution. Conversely, EDE models are constructed with specific dynamical mechanisms that make the energy density significant only during precise epochs, after which it disappears, being jeopardized by unpleasant fine-tuning issues.

\section{Experimental set up}\label{sezione3}

To prove the existence of our additional fluid with density $\Omega_s(1+z)^{3(1+\omega_s)}$, we define the underlying background cosmology as an extension of the standard $\Lambda$CDM framework, in which the comoving distance is defined as
\begin{equation}
\label{eq:dm}
D_{\rm M}(z) = c\int_0^{z}\frac{dz^\prime}{H(z^\prime)},
\end{equation}
where the parameters describing the expansion history enter in the definition of the Hubble parameter
\begin{equation}
H(z) = H_0 \sqrt{\Omega_\nu(z)+{\sum}_i \Omega_i (1+z)^{3(1+\omega_i)}},
\label{eq:hz}
\end{equation}
where the set, $i=\{\Lambda,m,\gamma,s\}$,  labels the cosmological constant, dust, radiation and \emph{matter with pressure}, respectively, exhibiting  barotropic indexes as $\omega_i=\{-1,0,1/3,\omega_s\}$.
The energy density of neutrinos $\Omega_\nu(z)$ scales like radiation (with EoS $\omega_\nu=1/3$)
at early times and like matter (with $\omega_\nu=0$) at late times.
Ensuring a spatially flat universe, $\Omega_k=0$, naturally implies $\Omega_\Lambda = 1 - \Omega_m - \Omega_s - \Omega_\gamma - \Omega_\nu$.
Thus, from Eq.~\eqref{eq:dm}, we can define the luminosity distance $D_{\rm L}(z)=(1+z)D_{\rm M}(z)$ and the diameter angular distance $D_{\rm A}(z) = D_{\rm M}(z)/(1+z)$.

\subsection{Numerical results}

We utilize the following data sets: the \textit{Pantheon+} sample of SNe Ia, the second data release (DR2) of baryonic acoustic oscillations (BAOs) from the Dark Energy Spectroscopic Instrument (DESI), and the CMB shift parameters, indicated in detail below and in Appendix~\ref{app:A}.

\begin{itemize}
\item[-] {\bf Pantheon+ and SH0ES.}
The Pantheon+ is a catalog of $1701$ SNe Ia with redshifts $z\in[0,2.3]$.
It comprises 18 different samples \citep{2022ApJ...938..113S} and includes the released
SH0ES Cepheid host distance anchors that provide the value of $H_0$ in tension with CMB determination  \cite{Brout:2022vxf}.

When using SH0ES Cepheid host distances, the $1701$ SN distance modulus residuals are given by
\begin{equation}
\label{eq:dmuprime}
\Delta \mu_i=
\begin{cases}
\mu_i - \mu_{{\rm C},i} & i \in \text{Cepheid hosts}, \\
\mu_i - \mu_{{\rm th}}(z_i) &\text{otherwise} ,
\end{cases}
\end{equation}
where $\mu_i$ are the inferred SN distance moduli, $\mu_{{\rm C},i}$ are the Cepheid-calibrated host-galaxy distance moduli by SH0ES, and $\mu_{\rm th}(z_i)$ are the model distance moduli defined as
\begin{equation}
\mu_{\rm th}(z_i) = m_i - M = 5\log(D_{\rm L}(z_i)/10\,{\rm pc}) ,
\end{equation}
in which $m_i$ are the rest-frame $B$-band apparent magnitudes of each SN Ia and $M$ is the rest-frame $B$-band absolute magnitude, that can be viewed as a nuisance parameter and marginalized \cite{SNLS:2011lii}. Thus, the cosmological parameters can be constrained by maximizing the loglikelihood
\begin{equation}
\label{eq:likelihood}
\ln \mathcal{L}_{\rm SN} = -\frac{1}{2} \Delta \vec{\mu}^T~C^{-1}~\Delta \vec{\mu} ,
\end{equation}
where $C=C_{\rm SN}+C_{\rm C}$ is the total covariance matrix of SNe and Cepheid host-distances (accounting for both statistical and systematic errors).\footnote{\url{https://github.com/PantheonPlusSH0ES/DataRelease}}
\item[-] {\bf DR2 DESI-BAO.} The DR2 of DESI-BAO data set is shown in Tab.~\ref{tab:DESIBAO}.
We utilize a total of $N_{\rm D} = 13$ measurements: $6$ transverse distances $D_{\rm M}$, $6$ Hubble rate distances $D_{\rm H}$ ($6$ data points), and one angle-averaged distance $D_{\rm V}$, all normalized over comoving sound horizon at the baryon drag redshift $z_d$, i.e., $r_d=r_s(z_d)$  \cite{DESI:2025zgx}.
To break the $r_d-H_0$ degeneracy is mandatory to provide the explicit expression of $r_d$ as a function of the model parameters given in Sect.~\ref{sezione2}. Here, we only fix $z_d = 1059.94 \pm 0.30$ \cite{Planck2018}.

Provided that $D_{\rm M}$ is given in Eq. \eqref{eq:dm}, the other two distance measurements are given by
\begin{subequations}\label{distances}
\begin{align}
\frac{D_{\rm H}(z)}{r_d}& = \frac{c}{r_d H(z)},\\
\frac{D_V(z)}{r_d} &= \frac{\left[z D_{\rm H}(z) D_{\rm M}^2(z) \right]^{1/3}}{r_d},
\end{align}
\end{subequations}
The total BAO log-likelihood is thus given by
\begin{equation}
    \ln\mathcal{L}_{\rm BAO}\propto -\frac{1}{2}\sum^{N_D}_{i=1}\left[\frac{Y_i-Y(z_i)}{\sigma_{Y_i}}\right]^2,
\end{equation}
where $Y_i$ and $\sigma_{Y_i}$ are DR2 data and errors, whereas $Y(z_i)= \{D_M(z_i)/r_d,
D_H(z_i)/r_d, D_V(z_i)/r_d\}$ are given in Eqs.~\eqref{eq:dm} and \eqref{distances}.
\item[-]{\bf CMB shift parameters.}
We employ the CMB shift parameters such as $\mathcal R=1.7502\pm 0.0046$ and $l_A=301.471^{+0.089}_{-0.090}$ \cite{Chen:2018dbv} that are related, particularly $l_A$, to the acoustic angular scale at the recombination, i.e., $\theta_\star = r_\star/D_{\rm M}(z_\star)$.
The shift parameters are defined as
\begin{subequations}
    \begin{align}
\label{eq:R}
\mathcal R(z_{*}) &= \frac{D_{\rm M}(z_\star) H_0 \sqrt{\Omega_m}}{c},\\
\label{eq:lA}
l_{\rm A}(z_{*}) &= \pi \frac{D_{\rm M}(z_\star)}{r_\star},
\end{align}
\end{subequations}
in which we can fix $z_\star= 1089.92 \pm 0.25$ \cite{Planck2018}, and take the explicit expression of $r_\star$ as a function of the model parameters given in Eqs.~\eqref{eq:0} and \eqref{eq:cszel}.

The total CMB log-likelihood is
\begin{equation}
    \ln\mathcal{L}_{\rm CMB}\propto -\frac{1}{2}\sum_{X}\left[\frac{X-X(z_\star)}{\sigma_{X}}\right]^2,
\end{equation}
where $X$ and $\sigma_{X}$ are $\mathcal R$ and $l_A$ with the associated errors, whereas $X(z_\star)= \{\mathcal R(z_\star),
l_A(z_\star)\}$ are given in Eqs.~\eqref{eq:R}--\eqref{eq:lA}.
\end{itemize}

By accounting for all the above described data sets, we can deduce the model cosmological parameters by maximizing the total log-likelihood function defined as
\begin{equation}
\ln\mathcal{L} = \ln \mathcal L_{\rm SN} + \ln \mathcal L_{\rm BAO} + \ln \mathcal L_{\rm CMB}.
\end{equation}

\begin{figure}
\includegraphics[width=\hsize,clip]{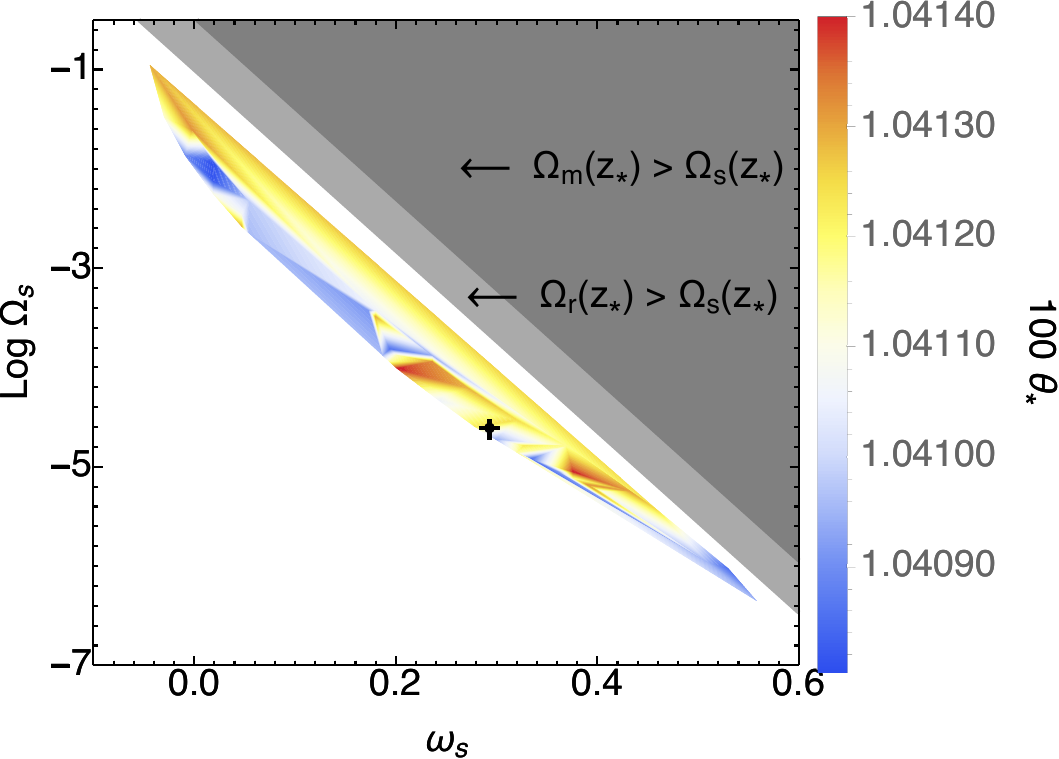}
\caption{Bounds in the $\omega_s-\log{\Omega_s}$ plane, compatible with the $\Lambda$CDM model constraints from SNe Ia, i.e., $\Omega_m = 0.334\pm0.018$, $H_0 = (73.6\pm1.1)$ km/s/Mpc, and from the CMB, i.e., $100\theta_\star=1.04110 \pm 0.00031$ (see the right side color-coded bar).
The excluded areas are the ones where $\Omega_r(z_\star)>\Omega_s(z_\star)$ (light gray) and $\Omega_m(z_\star)>\Omega_s(z_\star)$ (dark gray). The black dot with $2\sigma$ errors is the best-fit got from our MCMC simulations.}
\label{fig:Osws}
\end{figure}

Before proceeding with the MCMC fitting, it is useful to find the appropriate priors on the parameters $\Omega_s$ and $\omega_s$ of the additional fluid.
To this aim, we use the definitions from Sec.~\ref{sezione2} and the value of the acoustic angular scale $\theta_\star = 0.0104110 \pm 0.0000031$ \cite{Planck2018} introduced in Sec.~\ref{sezione3}.
Next, we use $\Omega_m = 0.334\pm0.018$, $H_0 = (73.6\pm1.1)$~km/s/Mpc inferred from SNe Ia, within the $\Lambda$CDM model \cite{2022ApJ...938..110B}.
In particular, the choice of the local estimate of $H_0$ gives us forecasts on the bounds of the additional fluid that might solve the Hubble tension.
Finally, we let $\Omega_s$ and $\omega_s$ vary with the condition that the deduced values of $\theta_\star$ shall stay within the observational constraints.
The results are shown in Fig.~\ref{fig:Osws}, where the permitted values in the $\omega_s-\log{\Omega_s}$ plane satisfy the conditions $\Omega_s(z_\star)<\Omega_r(z_\star)$ and $\Omega_s(z_\star)<\Omega_m(z_\star)$, where $\Omega_r(z_\star)=\Omega_\gamma(z_\star)+\Omega_\nu(z_\star)$ since at recombination neutrinos behave like radiation.

Therefore, the deduced priors are:
\begin{equation}
\nonumber
\begin{array}{rclcrclr}
H_0 & \in & \left[50,100\right]{\rm km/s/Mpc}, &\quad & \quad \Omega_m & \in &\left[0,1\right],\\
\log\Omega_s & \in & \left[-8,0\right], & \quad & \omega_s & \in & \left[-0.1,0.6\right].
\end{array}
\end{equation}

MCMC fitting is performed using \texttt{CLASS}, a highly modular, accurate and fast code that can be easily extended for custom physics and it is widely used in cosmology for fitting CMB, last scattering surface physics, and so on.
This code solves the Boltzmann equations for perturbations in the early universe and accounts for the following components: photons, baryons, cold dark matter, neutrinos (here the effective number of relativistic species is fixed to the standard value $N_{\rm eff}=3.044$), dark energy (under the form of a cosmological constant $\Lambda$ in our specific case), and optional extra fluids components (e.g., our matter fluid with pressure).
\begin{table}
\centering
\setlength{\tabcolsep}{.85em}
\renewcommand{\arraystretch}{1.3}
\begin{tabular}{lll}
\hline\hline
Parameter               & $\Lambda_{\omega_s}$CDM
                        & $\Lambda$CDM \\
\hline\hline
$H_0~({\rm km/s/Mpc})$  & $74.30^{+0.88(2.10)}_{-0.82(1.69)}$
                        & $68.70^{+0.45(0.79)}_{-0.36(0.87)}$  \\
$\Omega_m$              & $0.303^{+0.004(0.008)}_{-0.004(0.008)}$
                        & $0.294^{+0.003(0.007)}_{-0.004(0.007)}$\\
$\Omega_\Lambda$        & $0.697^{+0.004(0.008)}_{-0.004(0.008)}$
                        & $0.706^{+0.004(0.007)}_{-0.003(0.007)}$ \\
$\omega_s$              & $0.293^{+0.003(0.010)}_{-0.009(0.010)}$
                        & $-$ \\
$\Omega_s/10^{-5}$      & $2.48^{+0.32(0.58)}_{-0.28(0.60)}$
                        & $-$ \\
$\ln\mathcal L$         & $-654.86$
                        & $-671.60$\\
AIC (BIC)               & $1318\ (1340)$                                 & $1347\ (1358)$\\
$\Delta$AIC ($\Delta$BIC) & $0\ (0)$
                        & $29\ (18)$\\
\hline
\end{tabular}
\caption{Best-fit parameters with $1\sigma$ ($2\sigma$) errors and total log-likelihood values of $\Lambda_{\omega_s}$CDM and $\Lambda$CDM models, obtained by fitting SNe Ia, BAO, and CMB datasets. The last two rows list AIC (BIC) statistical criterion values and the corresponding differences.}
\label{tab:best}
\end{table}

Here, we performed the analysis using 50000 steps for our model, which yielded an acceptance rate of 0.51. In contrast, for the $\Lambda$CDM paradigm, we applied 40000 steps, achieving an acceptance rate of 0.64.
The results of the MCMC fit are summarized in Table~\ref{tab:best}.
In particular, the last two rows list the values of Aikake and Bayesian Information Criteria (AIC and BIC, respectively) \cite{2007MNRAS.377L..74L}
\begin{subequations}
\begin{align}
{\rm AIC}&=-2\ln \mathcal L + 2p\,,\\
{\rm BIC}&=-2\ln \mathcal L + p\ln N\,,
\end{align}
\end{subequations}
where $p$ is the number of parameters and $N=1716$ the whole sample size.
The $\Lambda_{\omega_s}$CDM model provides the lowest values of both AIC and BIC tests and, from the differences $\Delta{\rm AIC}\,(\Delta{\rm BIC})>6$, it is evident this model is strongly preferred over the $\Lambda$CDM paradigm.

\begin{figure}
\includegraphics[width=\hsize,clip]{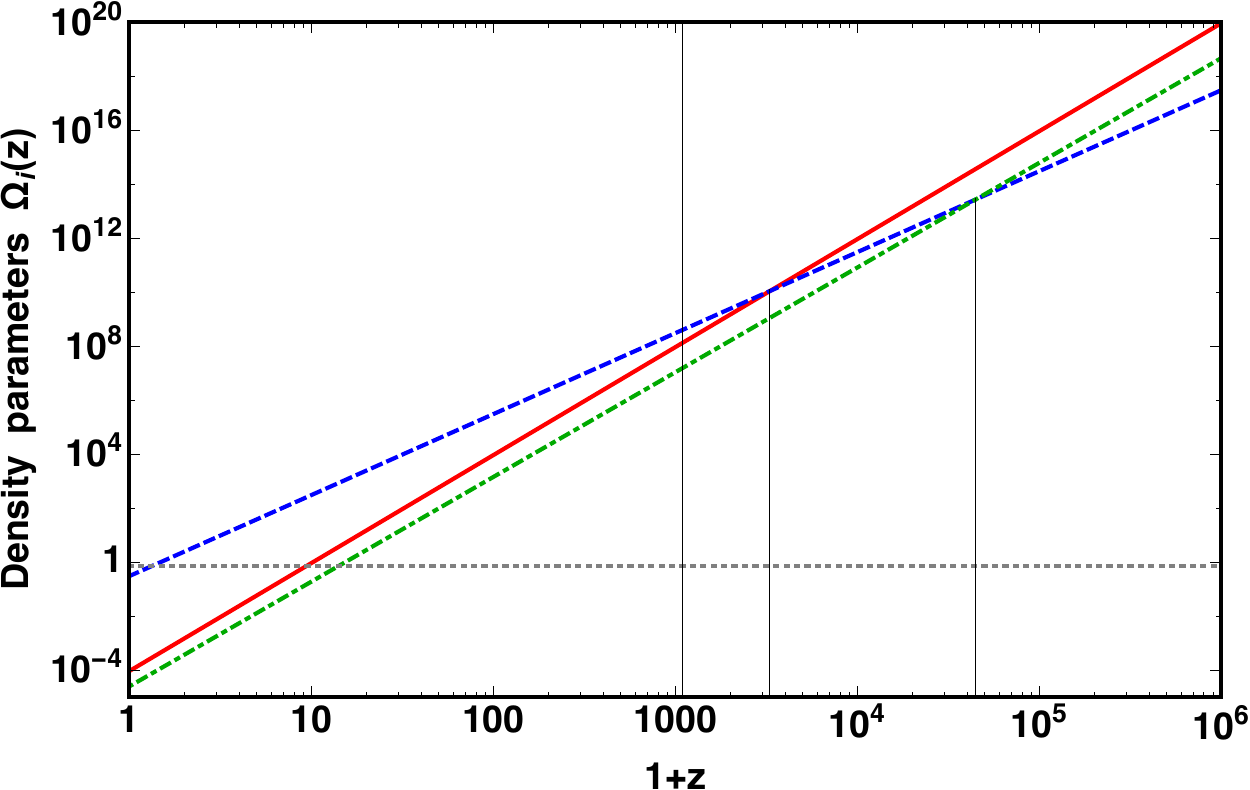}
\caption{Plot of the $\Lambda_{\omega_s}$CDM density parameters: $\Lambda$ (dotted gray), pressurelss matter (dashed blue), radiation (solid red), and matter with pressure (dot-dashed green). The vertical black lines (from left to right) mark the following redshifts: the recombination, pressureless matter--radiation equivalence, and pressureless matter--matter with pressure equivalence.}
\label{fig:eq}
\end{figure}

Using the results of the $\Lambda_{\omega_s}$CDM model from Table~\ref{tab:best}, we are now in the position to compute the equivalence epochs among fluids. Fig.~\ref{fig:eq} displays the density parameters $\Omega_i(z)=\Omega_i(1+z)^{3(1+\omega_i)}$ of all fluids.\footnote{In realizing this plot, for simplicity, we considered neutrinos like radiation at every redshifts, since when switching to matter their contribution to $\Omega_m$ is negligible.} In particular, we have marked with vertical black lines (from left to right): (i) the recombination redshift $z_\star$, (ii) the equivalence redshift between pressureless matter and radiation (photons and neutrinos) equivalence redshift ($z_{\rm eq}\approx3310$), and (iii) the equivalence redshift between matter with and without pressure ($z_{\rm ms}\approx44600$). From the plot it also evident that our matter fluid with pressure is: (a) subdominant with respect to pressureless matter and radiation (photons and neutrinos) at $z_\star$, and (b) subdominant with respect to photons at late times and photons+neutrinos at early times.
For the above reasons all the other model parameters are not significantly modified, with respect to the $\Lambda$CDM case, by the presence of such a further contribution.

The corresponding contour plots, obtained using \texttt{GetDist}, are portrayed in Figs.~\ref{fig:tri}--\ref{fig:tri1} of Appendix~\ref{app:B}.

\subsection{Physical consequences}

In our picture, ensuring an EoS forcedly positive and a Zel'dovic fluid implies a result
\begin{equation}
\omega_s=\frac{1}{3}-\epsilon,\qquad \Omega_s(z)=\Omega_s(1+z)^{4-3\epsilon},
\end{equation}
with $\epsilon=0.040^{+0.003}_{-0.009}$.
This kind of fluid, similar to radiation and here reinterpreted as matter with pressure, can be mainly attributed to specific cases of interest. Below, we report the main conclusions for each of them.

\begin{itemize}
    \item[-] {\bf Non-gauge fields like the Proca Lagrangian}. In such a case, the Hubble tension can be solved including the presence of massive electromagnetic fields, where the action of photon mass $m_{\rm P}$ reduces the EoS. For example, (a) ensuring the spatial component of the vector potential, ${\bf A}$, and the magnetic field, ${\bf B}$, to vanish, say ${\bf A}={\bf B}=0$, and (b) that for the scalar potential $A_0$ and the electric field $E$ holds the relation $m_{\rm P}^2 A_0^2 \ll E^2$, we end up with
    \begin{equation}
        \omega_{\rm P}\simeq \frac{1}{3}-\frac{4 A_0^2 m_{\rm P}^2}{3 E^2},\qquad \epsilon=\frac{4 A_0^2 m_{\rm P}^2}{3 E^2}.
    \end{equation}
    In a thermal background at temperature $T$, we have the following scalings: $A_0\approx T$ and $E^2\approx 2T^4$. Thus, using the above value of $\epsilon$ and the temperature at recombination, $T_\star=0.26$~eV, we obtain $m_{\rm P}=0.064^{+0.002}_{-0.007}$~eV, which is however excluded by the most conservative existing upper limit on the photon mass $m_\gamma<10^{-18}$~eV obtained from the solar wind magnetic field structure	\cite{Ryutov:2007zz}.

    \item[-] {\bf Dark photons}. Highly debated and recently under exam, dark photons are hypothetical massive gauge bosons associated with a hidden $U(1)_{\rm D}$ symmetry in a dark sector. They may have a small kinetic mixing with massless photons and very weak or no direct coupling to standard model matter \cite{Cline:2024qzv}.
    A massive dark photon behaves like a Proca field
    \begin{equation}
        \omega_{\rm D}\simeq \frac{1}{3}-\frac{4 A_0^2 m_{\rm D}^2}{3 E^2},\qquad \epsilon=\frac{4 A_0^2 m_{\rm D}^2}{3 E^2},
    \end{equation}
    but with much greater freedom, since dark photons are not subject to the same tight experimental bounds as standard photons.
    Therefore, for these constituents, we straightforwardly take the previous bound $m_{\rm D}=0.064^{+0.002}_{-0.007}$~eV.

    \item[-] {\bf Sterile neutrinos and thermal scalar fields}. Sterile neutrinos imply severe bounds on the effective number of relativistic species, whereas the case of scalar fields can be extended by considering the generalized K-essence described in Appendix~\ref{app:3}.
    Within the semi-relativistic approximation, we have
    \begin{equation}
        \omega_{\rm S}\simeq \frac{1}{3}-\frac{m_{\rm S}^2/3}{\kappa^2 T^2 +m_{\rm S}^2},\qquad \epsilon=\frac{m_{\rm S}^2/3}{\kappa^2 T^2 +m_{\rm S}^2},
    \end{equation}
    where $m_{\rm S}$ is the mass and $\kappa\approx2.7$ for neutrinos (from Fermi-Dirac statistic) and $\kappa\approx3.2$ for scalars (from Bose-Einstein statistic).
    Both massive components behave like radiation with $\omega_{\rm S}\approx 1/3$ when $m_{\rm S}\ll T$, and like pressureless matter with $\omega_{\rm S}\approx 0$ when $m_{\rm S}\gg T$. Limits on the mass, for instance at the recombination, can be obtained as follows
    \begin{equation}
        m_{\rm S}\simeq \kappa T_\star\sqrt{\frac{3\epsilon}{1-3\epsilon}}.
    \end{equation}
    Sterile neutrinos (fermions) shall be accounted for in $N_{\rm eff}$ and since in our MCMC fits it is fixed to the standard value $N_{\rm eff}=3.044$, any mass estimates would be physically incorrect. Conversely, for thermal scalar fields (bosons), there is no such a limitation, thus, we obtain  $m_{\rm S} = 0.31^{+0.01}_{-0.04}$~eV.

\end{itemize}

In view of our findings, our interpretation of mass with pressure lies on the fact that such a fluid may correspond to either a thermalized scalar field, whose nature can be associated with the quasi-quintessence, or to vector fields, in which the Proca field could represent an immediate and plausible explanation. Alternatives, making use of extended theories of gravity and/or modifications have not been faced, being beyond the scope of this work, see e.g., Refs.~\cite{vanderWesthuizen:2025iam,Paliathanasis:2025dcr,Paliathanasis:2025hjw}.

\section{Final remarks}\label{sezione4}

In this work, we focus on the early-time universe to heal the observational $H_0$ tension and proposed an alternative view to EDE models, namely \emph{invoking the existence of an additional barotropic fluid component}, physically inspired by stiff matter-like fluids, exhibiting relativistic properties differently from dust.

In so doing, we ensured this fluid to obey the Zel'dovic condition on the EoS, i.e., to possess $\omega_s>0$.
This matter-like constituent is constructed with non-negligible pressure, being quite distinct from both radiation and standard EDE candidates, as well as pure dust-like counterparts or neutrinos. Thus, by introducing such a component into the early cosmological dynamics, we directly altered the expansion rate $H(z)$ and the comoving sound horizon $r_s$ prior to recombination, modifying moreover the sound speed $c_s$. This enabled an effective increase in $H_0$, through a minimal and direct extension of the $\Lambda$CDM paradigm, here baptized $\Lambda_{\omega_s}$CDM model.

To prevent any significant modification of cosmic measurements, we required the additional barotropic component to remain subdominant with respect to  radiation, matter, and dark energy throughout the entire cosmic expansion history. To ensure this condition, we computed exclusion regions in the parameter space defined by the fluid’s energy density and EoS, demonstrating that its contribution remained significantly smaller than that of radiation and thus did not produce any substantial alteration in the CMB  anisotropy power spectrum.

Thus, once viable priors have been determined, we carried out a full cosmological parameter estimation using a MCMC simulation, involving Pantheon+ with SH0ES SNe, the DR2 of DESI-BAO, and the CMB shift parameters data sets. To this end, we modified the \texttt{CLASS} code \cite{2011arXiv1104.2932L} to include the dynamics of the extra fluid, and performed the analysis in accordance with the latest Planck observational constraints. In this respect, we included radiation and neutrino contributions, employing the assumption of spatial flatness.

We showed that the proposed extension is favored with respect to the  $\Lambda$CDM benchmark model, while \emph{our findings ruled out the case of a purely stiff fluid} with EoS $\omega = 1$, instead favoring a scenario in which the fluid exhibits an EoS slightly below that of photons $\omega_s \lesssim \omega_\gamma$, and a density parameter smaller than the photons background, being $\Omega_s/\Omega_\gamma\simeq 0.45$.

Afterwards, we discussed the physical implications of such a fluid, emphasizing its phenomenological consistency and  theoretical plausibility. We thus proposed that the fluid may be interpreted in view of a possible Proca-like field, where the effect of nonzero mass for the vector field can alter the EoS, departing from the genuine case given by radiation. We thus concluded that at least \emph{two vector fields may be present just before the last scattering surface}, without, however, excluding more exotic scenarios, involving dark radiation, axions, and so on.

In this respect, we also inferred constraints on equivalence and transition between
epochs between pressureless matter and radiation (occurring at a redshift $z_{\rm eq}\approx3310$), and between matter with and without pressure (occurring at a redshift $z_{\rm ms}\approx44600$). The matter fluid with pressure is subdominant with respect to pressureless matter and radiation at $z_\star$, and always subdominant with respect to radiation.
These features show that all other constituents are not significantly modified by the presence of such a further contribution.

Future works will shed light on comparing this matter with pressure fluid with other approaches that predict matter with pressure, by complicating the barotropic factors, or simply involving thermodynamic alternatives than a constant and positive $\omega_s$. Moreover, we will see more deeply the influence on structure formation of this fluid and, above all, its fundamental nature, thus exploring more deeply the role of Proca-like fields in early-time cosmology. A direct comparison with additional constituents, such as neutrinos, relativistic matter, etc. will also be the object of future speculations. Last but not least, further experimental settings will be pursued with the aim of refining our bounds over the free parameters of our fluid.

\section*{Acknowledgements}

OL acknowledges Maryam Azizinia for discussion and support by the  Fondazione  ICSC, Spoke 3 Astrophysics and Cosmos Observations National Recovery and Resilience Plan (Piano Nazionale di Ripresa e Resilienza, PNRR) Project ID $CN00000013$ ``Italian Research Center on  High-Performance Computing, Big Data and Quantum Computing" funded by MUR Missione 4 Componente 2 Investimento 1.4: Potenziamento strutture di ricerca e creazione di ``campioni nazionali di R\&S (M4C2-19 )" - Next Generation EU (NGEU). MM acknowledges
support by the project OASIS, “PNRR Bando a Cascata da INAF M4C2 - INV. 1.4”.

\appendix

\section{Data sets used in the analysis}\label{app:A}

Pantheon+ SN Ia and Cepheid host distance moduli, together with the covariance matrixes $C_{\rm SN}$ and $C_{\rm C}$ can be found at \url{https://github.com/PantheonPlusSH0ES/DataRelease}.

DR2 DESI-BAO, consisting of $N_{\rm D} = 13$ measured values of $D_M(z)/r_d$, $D_H(z)/r_d$, and $D_V(z)/r_d$ are taken from \citet{DESI:2025zgx} and shown in Tab.~\ref{tab:DESIBAO}.
No covariance matrix has been input, since not released yet.
\begin{table*}
\centering
\setlength{\tabcolsep}{2.5em}
\renewcommand{\arraystretch}{1.1}
\begin{tabular}{lcccc}
\hline\hline
Tracer     & $z_{\rm eff}$ & $D_{\rm M}/r_d$ & $D_{\rm H}/r_d$ & $D_{\rm V}/r_d$ \\
\hline
BGS & $0.295$ & $-$ & $-$ & $7.942\pm 0.075$  \\
LRG1 & $0.510$ & $13.588\pm 0.167$ & $21.863\pm 0.425$ & $-$  \\
LRG2 & $0.706$ & $17.351\pm 0.177$ & $19.455\pm 0.330$ & $-$  \\
LRG3+ELG1 & $0.934$ & $21.576\pm 0.152$ & $17.641\pm 0.193$ & $-$ \\
ELG2 & $1.321$ & $27.601\pm 0.318$ & $14.176\pm 0.221$ & $-$  \\
QSO & $1.484$ & $30.512\pm 0.760$ & $12.817\pm 0.516$ & $-$  \\
Lya QSO & $2.330$ & $38.988\pm 0.531$ & $8.632\pm 0.101$ & $-$ \\
\hline
\end{tabular}
\caption{The DR2 DESI data points with associated errors for bright galaxy survey (BGS), luminous red galaxies (LRG),
emission line galaxies (ELG), quasars (QSO), Lyman-$\alpha$ forest quasars (Lya QSO) and a combination of LRG+ELG \cite{DESI:2025zgx}.}
\label{tab:DESIBAO}
\end{table*}
The CMB shift parameters $\mathcal R$ and $l_A$ are taken from \citet{Chen:2018dbv}.

\section{Experimental results}\label{app:B}

\begin{figure*}
\includegraphics[width=0.7\hsize,clip]{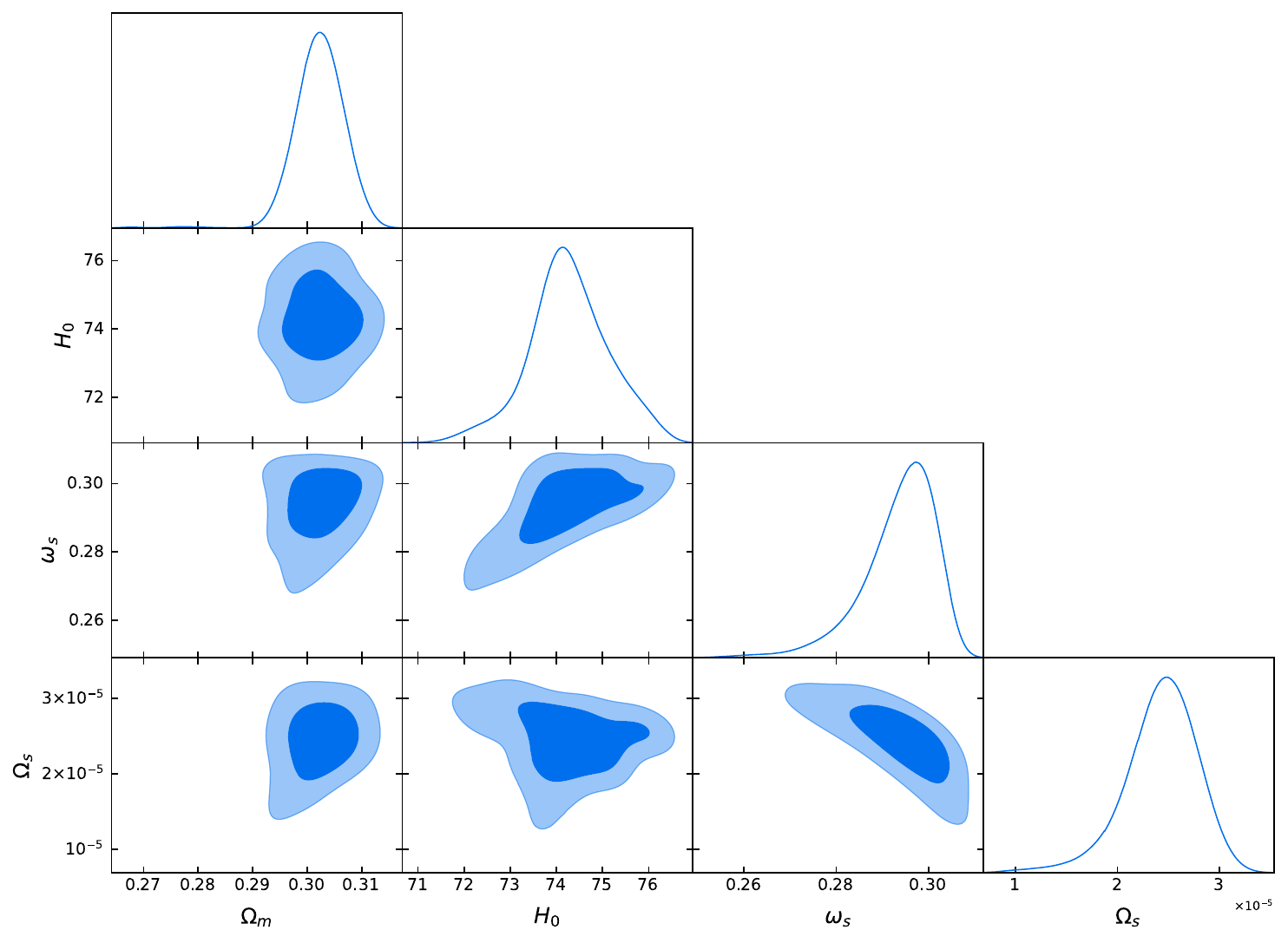}
\caption{Contour plots of our $\Lambda_{\omega_s}$CDM model with respect to the data points employed in this work. The contours have been obtained by fixing cosmic radiation, in a universe with a further fluid, constituted by matter, dark energy under the form of a cosmological constant, $\Lambda$, plus our barotropic fluid, whose density is given by $\Omega_s(z)=\Omega_s a^{-3(1+\omega_s)}$, ensuring $\omega_s>0$.}
\label{fig:tri}
\end{figure*}

The contour plots of the model parameters with best-fit values listed in Table~\ref{tab:best} are shown in Fig.~\ref{fig:tri}.

\begin{figure*}
\includegraphics[width=0.6\hsize,clip]{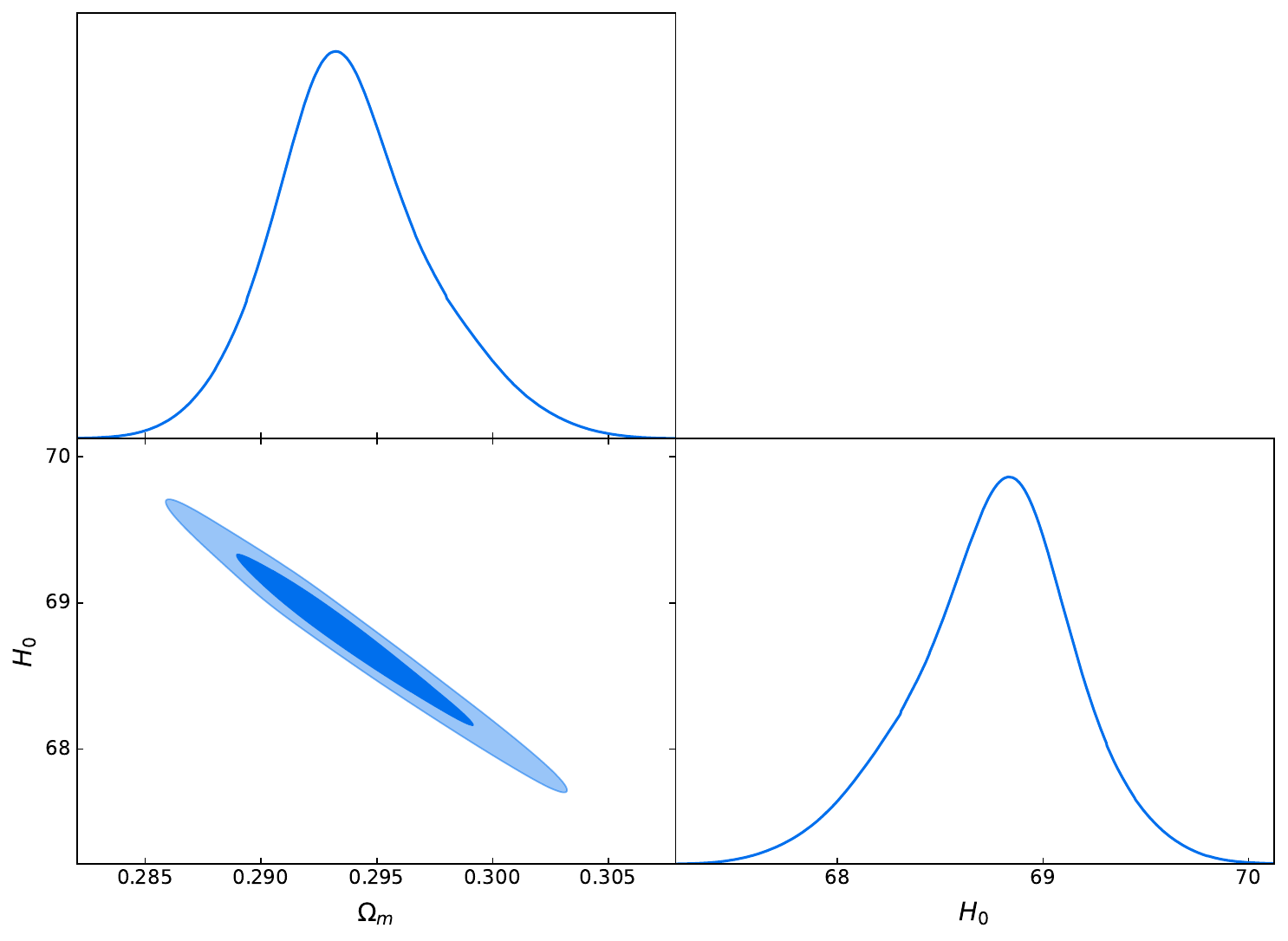}
\caption{Contour plots of the $\Lambda$CDM model with respect to the data points employed in this work.}
\label{fig:tri1}
\end{figure*}

\section{Alternative models of beyond-dust matter}\label{app:3}

It is possible to extend the EoS for matter, including pressure, by considering alternatives to the standard dust hypothesis. For example, barotropic fluids can be considered using effective unified dark energy-dark matter models \cite{Dunsby:2016lkw}, or working extensions of the Chaplygin gas that appear suitable at the level of background and perturbations \cite{Dunsby:2023qpb}. Alternatives, quite more complicated, would imply the presence of barotropic fluids that act differently from standard ones, namely profoundly modifying the EoS, including complicated terms inspired by solid state physics, such as logotropic fluids \cite{Boshkayev:2021uvk,Capozziello:2022ygp,Boshkayev:2019qcx}, Anton-Schmidt approaches \cite{Capozziello:2017buj,Capozziello:2018mds}, Murnagham EoS \cite{Dunsby:2024ntf} and so on.

\subsection{Generalized K-essence models}

Matter with pressure can be accounted involving a barotropic fluid with constant EoS \cite{Luongo:2022uva,Boshkayev:2020vrg}, albeit a fundamental representation of such a prerogative can be obtained ensuring $c_s=0$ within a scalar field Lagrangian, whose effective form can be described through a Lagrange multiplier as
\cite{2018PhRvD..98j3520L}, $\mathcal{L} = \lambda\, Y\left(X,\varphi\right) + K_0 - V\left(\varphi\right)$,
with the generalized kinetic term of the scalar field $X$, kinetic functions $K_0={\rm const}$ and $Y$, the Lagrangian multiplier $\lambda$, and the potential $V$, yielding both spontaneous symmetry breaking and chaotic inflation pictures \cite{2022CQGra..39s5014D}.

The above paradigm defines density and pressure
\begin{subequations}
\label{Pandrho}
\begin{align}
\rho =\, &2X \partial_X\mathcal{L} - \left(K_0 -  V \right)\,,\\
P =\, &K_0 - V\,.
\end{align}
\end{subequations}
Distinguishing between baryonic (\emph{b}) and cold dark matter (\emph{c}) yields $K_0=K_{b,0}+K_{c,0}$ and $\partial_X\mathcal L = \lambda \,\partial_X\!\left(Y_b+Y_c\right)$.

After the transition induced by the symmetry breaking, the field $\varphi$ reaches the minimum of the potential $V_0=-0.22\,M_{\rm P}^4$, where $M_{\rm P}$ is the Planck mass, and the magnitude of dark matter pressure $K_{c,0}\gg K_{b,0}$ can be adjusted to cancel out the vacuum energy density, healing the fine tuning issue associated to the cosmological constant $\Lambda$ \cite{Belfiglio:2022qai,Belfiglio:2023rxb}. In this picture, Eqs.~\eqref{Pandrho} become
\begin{subequations}
\label{PandrhoA}
\begin{align}
\rho &\approx 2X\lambda \,\partial_X\!\left(Y_b+Y_c\right) - K_{b,0}\,,\\
P &\approx  K_{b,0}\,.
\end{align}
\end{subequations}
The above fluid is isentropic, leading to the identification $2X\lambda \,\partial_X\!\left(Y_b+Y_c\right) \equiv \left(\rho_b + \rho_{c}\right)(1+z)^3$ that describes the densities of pressureless dark matter and baryonic matter endowed with negative pressure, $K_{b,0}$. For additional details, see Ref.~\cite{2018PhRvD..98j3520L}.

\subsection{Anton-Schmidt and logotropic fluids}

The issue of negative pressure poses a problem to obtain direct experimental evidence of negative pressure in laboratory. However, in condensed matter physics, it is possible to argue a pressure that turns out to be locally negative. A notable example is given by the Anton–Schmidt EoS \cite{AntonSchmidt1997}, that empirically describes the pressure of a crystalline solid under isotropic deformation. For cosmological applications, it becomes
\begin{equation}\label{anton}
    P = A\, \tilde{\rho}^{\gamma_G+\frac{1}{6}} \ln \tilde{\rho},
\end{equation}
where $\gamma_G$ denotes the so-called Grüneisen parameter, whose macroscopic definition is related to the thermodynamic properties of the material.
Here, $\tilde{\rho} \equiv \rho/\bar{\rho}$ is the normalized density with respect to a reference density $\bar{\rho}$, and $A$ represents the amplitude of the fluid.

Remarkably, as $\gamma_G \to -1/6$, the fluid becomes purely logotropic, i.e., described by
\begin{equation}\label{logo}
    P_l = A \ln \tilde{\rho}_l,
\end{equation}
initially proposed in Refs.~\cite{2015EPJP..130..130C,2016PhLB..758...59C}, where the subscript $l$ indicates its logotropic nature. The model has been motivated by the hypothesis that the same fluid acting as dark energy could also account for dark matter in galaxies, thus alleviating the cusp problem observed at the centers of spiral galaxies \cite{2015EPJP..130..130C}.

These two examples of matter with pressure, represented by Eqs.~\eqref{anton}--\eqref{logo}, also constitute possible unified dark energy models, albeit their validities have been severely criticized see e.g. \cite{Carloni:2024zpl,2021PhRvD.104b3520B}.

\subsection{The Chaplygin and Murnaghan fluids }

The Chaplygin gas was originally introduced as a simple prototype of unified dark energy models. This is achieved through an EoS that scales inversely with the density itself. However, such models are ruled out by observational data \cite{Fabris:2010vd,Carloni:2024zpl}.

More recently, motivated by a two-fluid interacting Lagrangian framework, where one fluid transports vacuum energy, it has been proposed that the pressure can be expressed by the Murnaghan EoS
\begin{equation}\label{P1}
    P = -A \left(\tilde\rho^{-\alpha}-1\right)\,,
\end{equation}
that corresponds to a Chaplygin gas supplemented by a cosmological constant term, where $A$ is the amplitude of the fluid with density normalized as $\tilde{\rho} \equiv \rho/\bar{\rho}$, and $\alpha$ is a constant. This paradigm was first introduced in Ref.~\cite{Dunsby:2023qpb} and subsequently reviewed in Ref.~\cite{Dunsby:2024ntf}, demonstrating promising fits to cosmological observations.

Within all the above alternatives, it becomes clear that matter with pressure can be straightforwardly modeled by a constant EoS parameter, as realized in the two simplest cases, i.e., the generalized K-essence and stiff-like fluids. Hence, rather than complicating the CMB consequences adopting logotropic, Anton-Schmidt and/or Chaplygin gas, we focused above on our $\Lambda_{\omega_s}$CDM framework that suggests matter: 1) with enhanced elastic properties, 2) not fully-cold, 3) distinct from neutrinos, while closely approximating radiation.

\end{document}